# Mapping the Gnutella Network:
# Properties of Large-Scale Peer-to-Peer Systems and Implications for System Design

Matei Ripeanu [1] matei@cs.uchicago.edu
Ian Foster [1,2] foster@mcs.anl.gov
Adriana Iamnitchi [1] anda@cs.uchicago.edu

[1] Computer Science Department, The University of Chicago, 1100 E. 58th St., Chicago, IL 60637.
[2] Mathematics and Computer Science Division, Argonne National Laboratory, 9700 S. Cass Ave., MCS/211, Argonne, IL 60439.

***Abstract***

Despite recent excitement generated by the peer-to-peer (P2P) paradigm and the surprisingly rapid deployment of some P2P applications, there are few quantitative evaluations of P2P systems behavior. The open architecture, achieved scale, and self-organizing structure of the Gnutella network make it an interesting P2P architecture to study. Like most other P2P applications, Gnutella builds, at the application level, a virtual network with its own routing mechanisms. The topology of this virtual network and the routing mechanisms used have a significant influence on application properties such as performance, reliability, and scalability. We have built a "*crawler*" to extract the topology of Gnutella's application level network. In this paper we analyze the topology graph and evaluate generated network traffic. Our two major findings are that: (1) although Gnutella is not a pure power-law network, its current configuration has the benefits and drawbacks of a power-law structure, and (2) the Gnutella virtual network topology does not match well the underlying Internet topology, hence leading to ineffective use of the physical networking infrastructure. These findings guide us to propose changes to the Gnutella protocol and implementations that may bring significant performance and scalability improvements. We believe that our findings as well as our measurement and analysis techniques have broad applicability to P2P systems and provide unique insights into P2P system design tradeoffs.

## 1. Introduction

Peer-to-peer systems (P2P) have emerged as a significant social and technical phenomenon. These systems provide infrastructure for communities that share CPU cycles (e.g., SETI@Home, Entropia) and/or storage space (e.g., Napster, FreeNet, Gnutella), or that support collaborative environments (Groove). Two factors have fostered the recent explosive growth of such systems: first, the low cost and high availability of large numbers of computing and storage resources, and second, increased network connectivity. As these trends continue, the P2P paradigm is bound to become more popular.

Unlike traditional distributed systems, P2P networks aim to aggregate large numbers of computers that join and leave the network frequently and that might not have permanent network (IP) addresses. In pure P2P systems, individual computers communicate directly with each other and share information and resources without using dedicated servers. A common characteristic of this new breed of systems is that they build, at the application level, a virtual network with its own routing mechanisms. The topology of the virtual network and the routing mechanisms used have a significant impact on application properties such as performance, reliability, scalability, and, in some cases, anonymity. The virtual topology also determines the communication costs associated with running the P2P application, both at individual hosts and in the aggregate. Note that the decentralized nature of pure P2P systems means that these properties are emergent properties, determined by entirely local decisions made by individual resources, based only on local information: we are dealing with a self-organized network of independent entities.

These considerations have motivated us to conduct a detailed study of the topology and protocols of a popular P2P system: Gnutella. In this study, we benefited from Gnutella's large existing user base and open architecture, and, in effect, used the public Gnutella network as a large-scale, if uncontrolled, testbed. We proceeded as follows. First, we captured the network topology, its generated traffic, and dynamic behavior. Then, we used this raw data to perform a macroscopic analysis of the network, to evaluate costs and benefits of the P2P approach, and to investigate possible improvements that would allow better scaling and increased reliability.

Our measurements and analysis of the Gnutella network are driven by two primary questions. The first concerns its connectivity structure. Recent research [1,8,7] shows that networks as diverse as natural networks formed by molecules in a cell, networks of people in a social group, or the Internet, organize themselves so that most nodes have few links while a tiny number of nodes, called hubs, have a large number of links. [14] finds that networks following this organizational pattern (power-law networks) display an unexpected degree of robustness: the ability of their nodes to communicate is unaffected even by extremely high failure rates. However, error tolerance comes at a high price: these networks are vulnerable to attacks, i.e., to the selection and removal of a few nodes that provide most of the network's connectivity. We show that, although Gnutella is not a pure power-law network, it preserves good fault tolerance characteristics while being less dependent than a pure power-law network on highly connected nodes that are easy to single out (and attack).

The second question concerns how well (if at all) the Gnutella virtual network topology maps to the physical Internet infrastructure. There are two reasons for analyzing this issue. First, it is a question of crucial importance for Internet Service Providers (ISP): if the virtual topology does not follow the physical infrastructure, then the additional stress on the infrastructure and, consequently, the costs for ISPs, are immense. This point has been raised on various occasions [9,12] but, as far as we know, we are the first to provide a quantitative evaluation on P2P application and Internet topology (mis)match. Second, the scalability of any P2P application is ultimately determined by its efficient use of underlying resources.

We are not the first to instrument and measure the Gnutella network. The Distributed Search Solutions (DSS) group has published results of their Gnutella surveys [4] and the Snowtella project [5] has focused on analyzing the characteristics of participating resources. Others have used this data to analyze Gnutella users' behavior [2], to analyze search protocols for power-law networks [6], and to forecast network growth through simulations [15]. However, our network crawling and analysis technology (developed independently of this work) goes significantly further in terms of scale (both spatial and temporal) and sophistication. While DSS presents only raw facts about the network, we analyze the generated network traffic, find patterns in network organization, and investigate its efficiency in using the underlying network infrastructure.

The rest of this paper is structured as follows. The next section succinctly describes Gnutella protocol and application. Section 3 introduces the crawler we developed to discover Gnutella's virtual network topology. In Section 4 we analyze the network and answer the questions introduced in the previous paragraphs. We conclude in Section 5.

## 2. Gnutella Protocol: Design Goals and Description

The Gnutella protocol [3] is an open, decentralized group membership and search protocol, mainly used for file sharing. The term Gnutella also designates the virtual network of Internet accessible hosts running Gnutella-speaking applications (this is the "Gnutella network" we measure) and a number of smaller, and often private, disconnected networks.

Like most P2P file sharing applications, Gnutella was designed to meet the following goals:

- *Ability to operate in a dynamic environment*. P2P applications operate in dynamic environments, where hosts may join or leave the network frequently. They must achieve flexibility in order to keep operating transparently despite a constantly changing set of resources.
- *Performance and Scalability*. The P2P paradigm shows its full potential only on large-scale deployments where the limits of the traditional client/server paradigm become obvious. Moreover, scalability is important as P2P applications exhibit what economists call the "network effect" [10]: the value of a network to an individual user scales with the total number of participants. Ideally, when increasing the number of nodes, aggregate storage space and file availability should grow linearly, response time should remain constant, while search throughput should remain high or grow.
- *Reliability*. External attacks should not cause significant data or performance loss.
- *Anonymity*. Anonymity is valued as a means of protecting the privacy of people seeking or providing unpopular information.

Gnutella nodes, called *servents* by developers, perform tasks normally associated with both SERVers and cliENTS. They provide client-side interfaces through which users can issue queries and view search results, accept queries from other servents, check for matches against their local data set, and respond with corresponding results. These nodes are also responsible for managing the background traffic that spreads the information used to maintain network integrity.

In order to join the system a new node/servent initially connects to one of several known hosts that are almost always available (e.g., gnutellahosts.com). Once attached to the network (e.g., having one or more open connections with nodes already in the network), nodes send messages to interact with each other. Messages can be *broadcasted* (i.e., sent to all nodes with which the sender has open TCP connections) or simply *back-propagated* (i.e., sent on a specific connection on the reverse of the path taken by an initial, broadcasted, message). Several features of the protocol facilitate this broadcast/back-propagation mechanism. First, each message has a randomly generated identifier. Second, each node keeps a short memory of the recently routed messages, used to prevent re-broadcasting and to implement back-propagation. Third, messages are flagged with time-to-live (TTL) and "hops passed" fields.

The messages allowed in the network are:

- *Group Membership* (PING and PONG) *Messages*. A node joining the network initiates a broadcasted PING message to announce its presence. When a node receives a PING message it forwards it to its neighbors and initiates a back-propagated PONG message. The PONG message contains information about the node such as its IP address and the number and size of shared files.
- *Search* (QUERY and QUERY RESPONSE) *Messages*. QUERY messages contain a user specified search string that each receiving node matches against locally stored file names. QUERY messages are broadcasted. QUERY RESPONSES are back-propagated replies to QUERY messages and include information necessary to download a file.

- *File Transfer* (GET and PUSH) *Messages*. File downloads are done directly between two peers using GET/PUSH messages.

To summarize: to become a member of the network, a *servent* (node) has to open one or many connections with nodes that are already in the network. In the dynamic environment where Gnutella operates, nodes often join and leave and network connections are unreliable. To cope with this environment, after joining the network, a node periodically PINGs its neighbors to discover other participating nodes. Using this information, a disconnected node can always reconnect to the network. Nodes decide where to connect in the network based only on local information; thus, the entire network form a dynamic, self-organizing network of independent entities. This virtual, application-level network has Gnutella servents at its nodes and open TCP connections as its links. In the following sections we present the techniques that we have developed to discover this network topology and analyze its characteristics.

## 3. Data Collection: The Crawler

We have developed a *crawler* that joins the network as a servent and uses the membership protocol (the PING-PONG mechanism) to collect topology information. In this section we briefly describe the crawler and discuss other issues related to data collection.

The crawler starts with a list of nodes, initiates a TCP connection to each node in the list, sends a generic join-in message (PING), and discovers the neighbors of the contacted node based on the PONG messages that it receives in reply. () Newly discovered neighbors are added to the list. For each discovered node the crawler stores its IP address, port, and the number of files and the total space shared. We started with a short, publicly available list of initial nodes, but over time we have incrementally built our own list with more than 400,000 nodes that have been active at one time or another.

We first developed a sequential version of the crawler. Using empirically determined optimal values for connection establishment timeout as well as for connection listening timeout (the time interval the crawler waits to receive PONGs after it has sent a PING), a sequential crawl of the network proved slow: about 50 hours even for a small network (4000 nodes). This slow search speed has two disadvantages: not only it is not scalable, but the dynamic network behavior means that the results obtained are far from a network topology snapshot.

In order to reduce the crawling time, we next developed a distributed crawling strategy. Our distributed crawler has a client/server architecture: the server is responsible with managing the list of nodes to be contacted, assembling the final graph, and assigning work to clients. Clients receive a small list of initial points and discover the network topology around these points. Although we could use a large number of clients (easily in the order of hundreds), we decided to use only up to 50 clients in order to reduce the invasiveness of our search. These techniques have allowed us to reduce the crawling time to a couple of hours even for a large list of starting points and a discovered topology graph with more than 30,000 active nodes.

Note that in the following we use a conservative definition of network membership: we exclude the nodes that, although reported as part of the network, our crawler could not connect to. This situation might occur when the local servent is configured to allow only a limited number of TCP connections or when the node leaves the network before the crawler contacts it.

## 4. Gnutella Network Analysis

We first summarize Gnutella network growth trends and dynamic behavior (Section 4.1). Our data gathered over a six month period show that although Gnutella overhead traffic has recently been decreasing, the generated traffic volume currently represents a significant percentage of total Internet traffic and is a major obstacle to further growth (Section 4.2). We continue with a macroscopic analysis of the network: we study first connectivity patterns (Section 4.3) and then the mapping of the Gnutella topology to the underlying networking infrastructure (Section 4.4).

### *4.1 Growth Trends and Dynamic Behavior*

Figure 1 presents the growth of the Gnutella network during a 6-month period. We ran our crawler during November 2000, February/March 2001, and May 2001. While in November 2000 the largest connected component of the network we found had 2,063 hosts, this grew to 14,949 hosts in March and 48,195 hosts in May 2001. Although Gnutella's failure to scale has been predicted time and again, the number of nodes in the largest network component grew about 25 times (admittedly from a low base) in a 6-month interval. It is worth mentioning that the number of connected components is relatively small: the largest connected component includes more than 95% of the active nodes discovered.

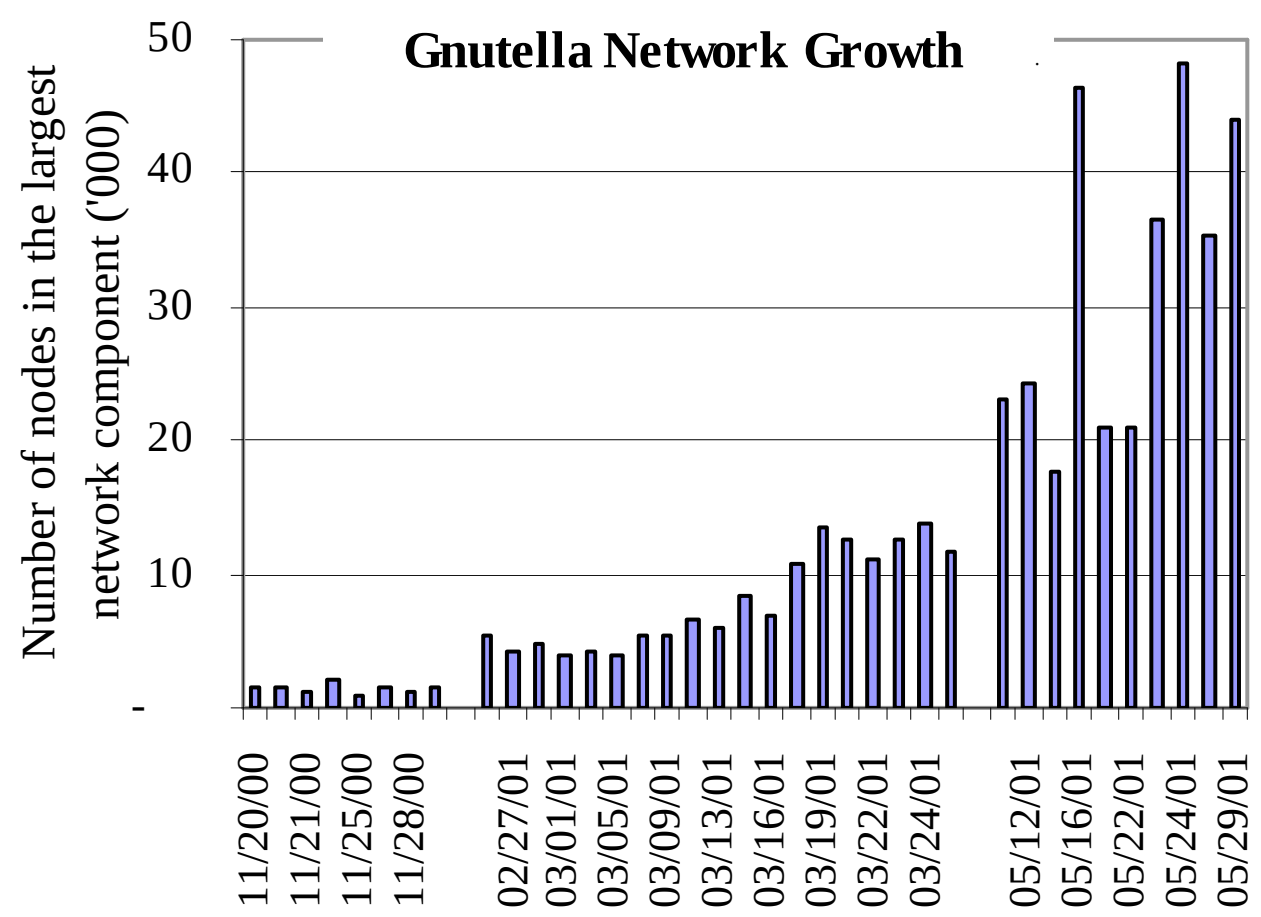

**Figure 1**: Gnutella network growth. The plot presents the number of nodes in the largest connected component in the network. Data collected during Nov. 2000, Feb./March 2001 and May 2001. We found a significantly larger network around Memorial Day (May 24-28) and Thanksgiving 2000, when apparently more people hunt for shared music online.

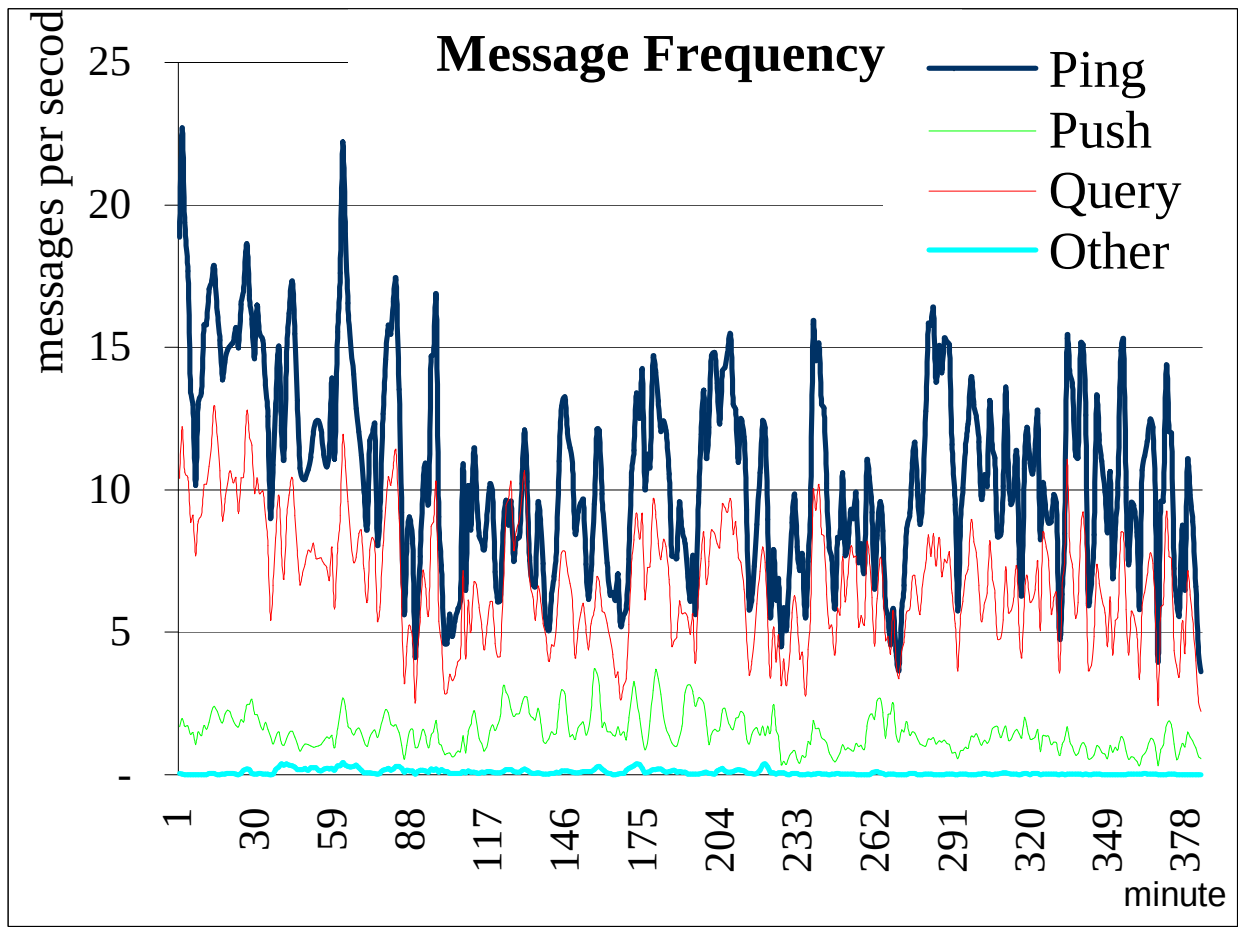

**Figure 2:** Generated traffic (messages/sec) in Nov. 2000 classified by message type over a 376 minute period. Note that overhead traffic (PING messages, that serve only to maintain network connectivity) formed more than 50% of the traffic. The only 'true' user traffic is QUERY messages. Overhead traffic has decreased by May 2001 to less than 10% of all generated traffic.

Using records of successive crawls, we investigate the dynamic graph structure over time. We discover that about 40% of the nodes leave the network in less than 4 hours, while only 25% of the nodes are alive for more than 24 hours. Given this dynamic behavior, it is important to find the appropriate tradeoff between discovery time and invasiveness of our crawler. Increasing the number of parallel crawling tasks reduces discovery time but increases the burden on the application. Obviously, the Gnutella map our crawler produces is not an exact 'snapshot' of the network. However, we argue that the network graph we obtain is close to a snapshot in a statistical sense: all properties of the network: size, diameter, average connectivity, and connectivity distribution are preserved.

### *4.2 Estimate of Gnutella Generated Traffic*

We used a modified version of the crawler to eavesdrop the traffic generated by the network. In Figure 2 we classify, according to message type, the traffic that goes across one randomly chosen link in November 2000. After adjusting for message size, we find that, on average, only 36% of the total traffic (in bytes) is user-generated traffic (QUERY messages). The rest is overhead traffic: 55% used to maintain group membership (PING and PONG messages) while 9% contains either non-standard messages (1%) or PUSH messages broadcast by servents that are not compliant with the latest version of the protocol. Apparently, after June 2001, these engineering problems were solved with the arrival of newer Gnutella implementations: generated traffic contains 92% QUERY messages, 8% PING messages and insignificant levels of other message types.

In Figure 3 we present the distribution of node-to-node shortest path length (the shortest path, in terms of number of links traversed, a message has to travel in order to get from one node to the other). Given that 95% of any two nodes are less than 7 hops away, the message time-to-live (TTL=7) preponderantly used, and the flooding-based routing algorithm employed, most links support similar traffic. We verified this theoretical conclusion by measuring the traffic at multiple, randomly chosen, nodes. As a result, the total Gnutella generated traffic is proportional to the number of connections in the network. Based on our measurements we estimate the total traffic (excluding file transfers) for a large Gnutella network as 1 Gbps: 170,000 connections for a 50,000-nodes Gnutella network times 6 Kbps per connection, or about 330 TB/month. To put this traffic volume into perspective we note that it amounts to about 1.7% of total traffic in US Internet backbones in December 2000 (as reported in [16]). We infer that the volume of generated traffic is an important obstacle for further growth and that efficient use of underlying network infrastructure is crucial for better scaling and wider deployment.

One interesting feature of the network is that, over a seven-month period, with the network scaling up almost two orders of magnitude, the average number of connections per node remained constant (Figure 4). Assuming this invariant holds, it is possible to estimate the generated traffic for larger networks and find scalability limits based on available bandwidth.

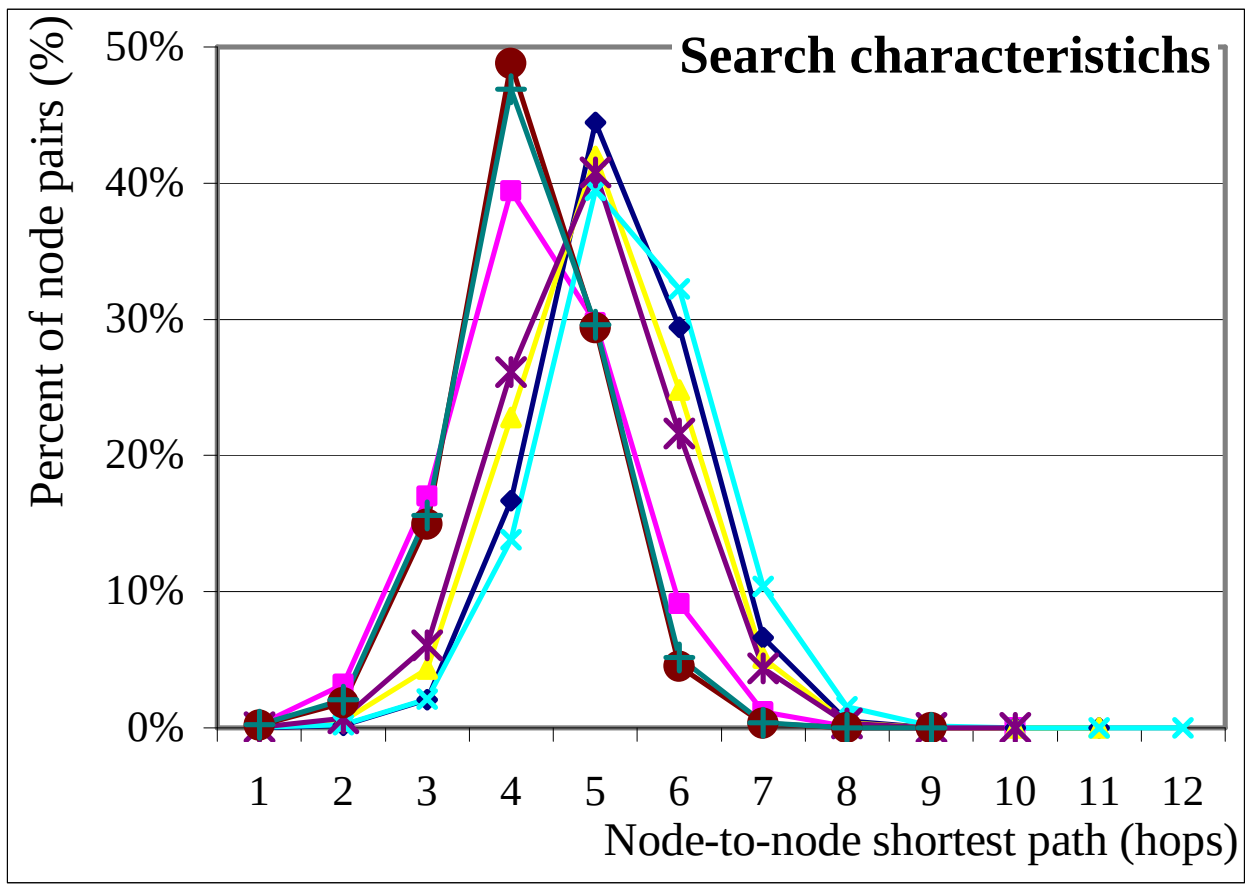

**Figure 3**: Distribution of node-to-node shortest paths. Each line represents one Gnutella network crawl. Note that, although the largest network diameter (the longest node-to-node path) is 12, more than 95% of node pairs are at most 7 hops away

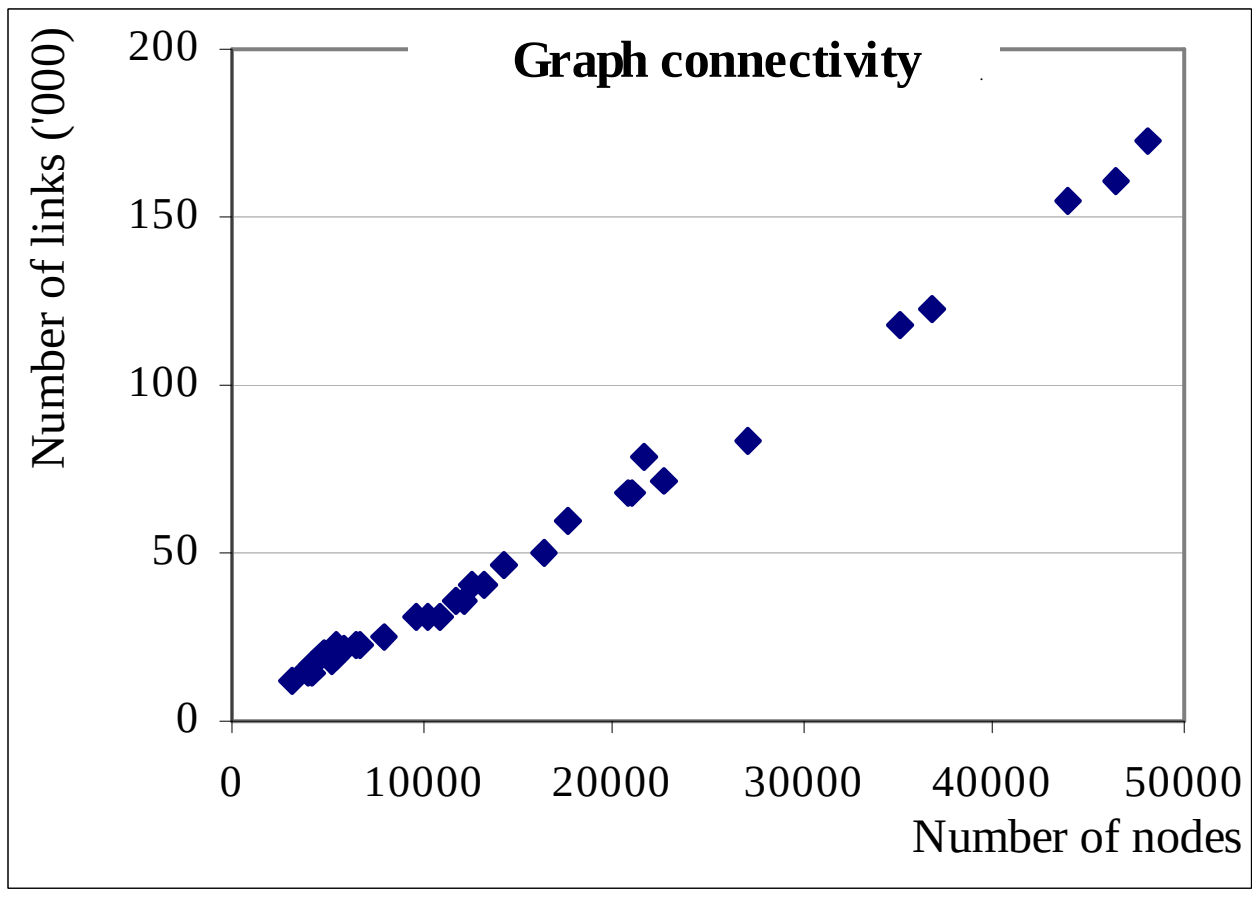

**Figure 4**: Average node connectivity. Each point represents one Gnutella network crawl. Note that, as the network grows, the average number of connections per node remains constant (average node connectivity is 3.4 connections per node).

### *4.3. Connectivity and Reliability in Gnutella Network. Power-law Distributions.*

When analyzing global connectivity and reliability patterns in the Gnutella network, it is important to keep in mind the self-organized network behavior: users decide only the maximum number of connections a node should support, and nodes decide whom to connect to or when to drop/add a connection based only on local information.

Recent research [1,7,8,13] shows that many natural networks such as molecules in a cell, species in an ecosystem, and people in a social group organize themselves as so called *power-law networks*. In these networks most nodes have few links and a tiny number of hubs have a large number of links. More specifically, in a power-law network the fraction of nodes with L links is proportional to $L^{-k}$, where $k$ is a network dependent constant.

This structure helps explain why networks ranging from metabolisms to ecosystems to the Internet are generally highly stable and resilient, yet prone to occasional catastrophic collapse [14]. Since most nodes (molecules, Internet routers, Gnutella servents) are sparsely connected, little depends on them: a large fraction can be taken away and the network stays connected. But, if just a few highly connected nodes are eliminated, the whole system could crash. One implication is that these networks are extremely robust when facing random node failures, but vulnerable to well-planned attacks.

Given the diversity of networks that exhibit power-law structure and their properties we were interested to determine whether Gnutella falls into the same category. Figure 5 presents the connectivity distribution in Nov. 2000. Although data are noisy (due to the small size of the networks), we can easily recognize the signature of a power-law distribution: the connectivity distribution appears as a line on a log-log plot. [6,4] confirm that early Gnutella networks were power-law. Later measurements (Figure 6) however, show that more recent networks tend to move away from this organization: there are too few nodes with low connectivity to form a pure power-law network. In these networks the power-law distribution is preserved for nodes with more than 10 links while nodes with fewer link follow an almost constant distribution.

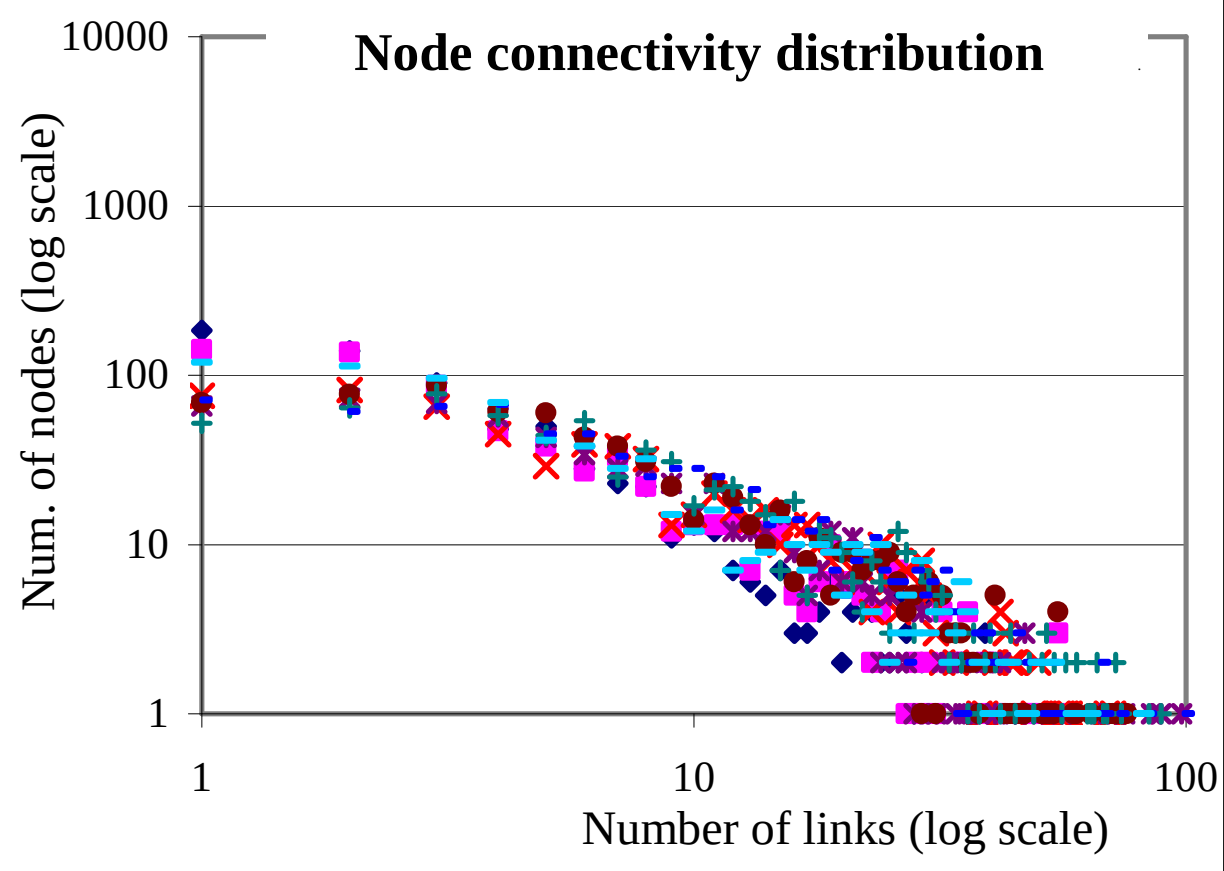

**Figure 5:** Connectivity distribution during November 2000. Each series of points represents one Gnutella network topology we discovered at different times during that month. Note the log scale on both axes. Gnutella nodes organized themselves into a power-law network.

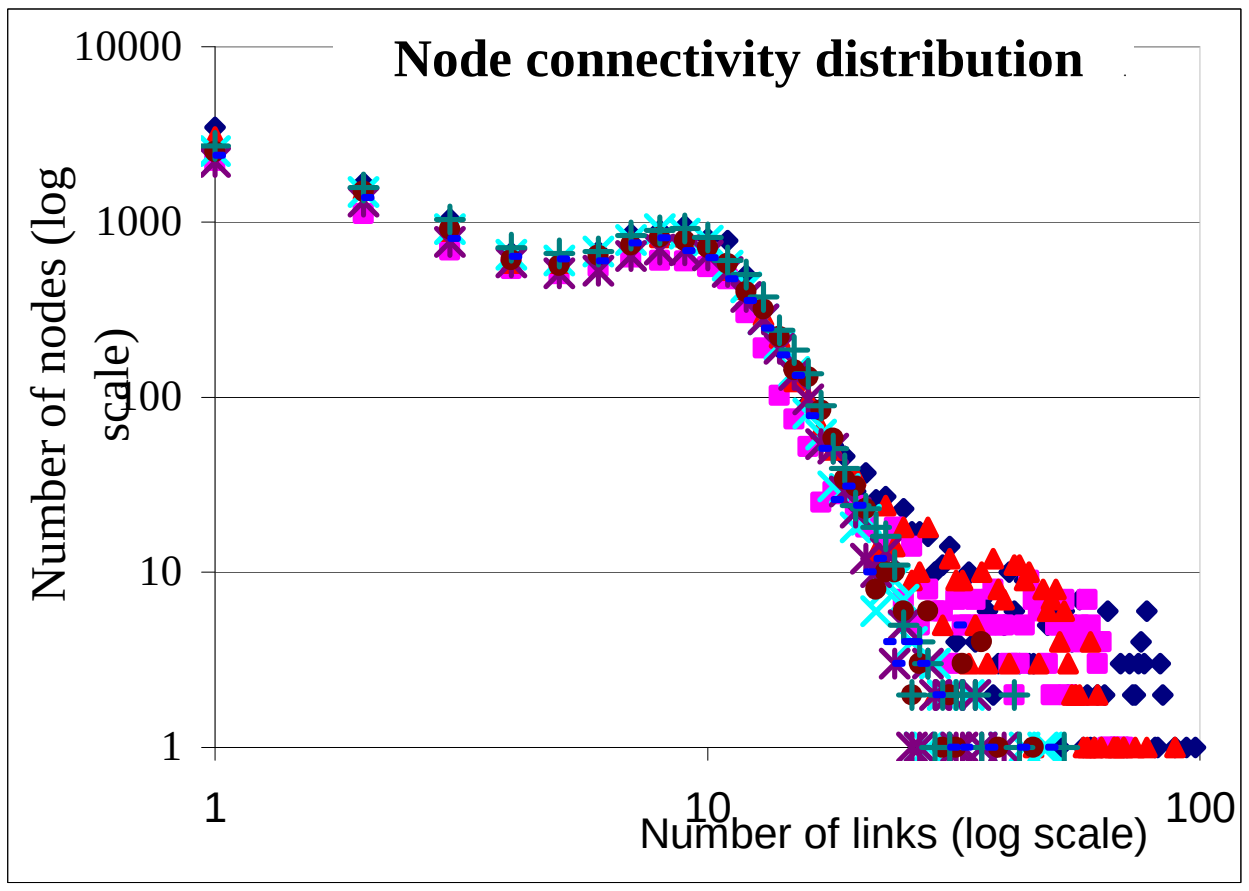

**Figure 6:** Connectivity distributions during March 2001. Each series of points represents one Gnutella network topology discovered during March 2001. Note the log scale on both axes. Networks crawled during May/June 2001 show a similar pattern.

We speculate there are two reasons for the peculiar distribution in Figure 6. First, Gnutella users are technically savvy users, early technology adopters. The percentage of Gnutella users with modem connection is significantly lower than in the Internet users population: less than 20% users connect through up to 100Kbps [5]. Moreover, the distribution of machines with better Internet connectivity (DSL and up) follows a power-law itself. Second, the lack of incentives to limit the traffic and the users perception that “more connections = better query results” leads users to employ as many connections as their network supports.

An interesting issue is the impact of this new, multi-modal distribution on network reliability. We believe that the more uniform connectivity distribution preserves the network’s ability to deal with random node failures while reducing the network dependence on highly connected, easy to single out (and attack) nodes.

We speculate that a group of devoted users maintain the small number of Gnutella nodes with the server-like characteristics visible in these power-law distributions. These nodes have a large number of open connections and/or provide much of the content available in the network. Moreover, these server-like nodes have a higher availability: they are about 50% more likely than the average to be found alive during two successive crawls.

### *4.4. Internet Infrastructure and Gnutella Network*

Peer-to-peer computing brings an important change in the way we use the Internet: it enables computers sitting at the edges of the network to act as both clients and servers. As a result, P2P applications change radically the amount of bandwidth consumed by the average Internet user. Most Internet Service Providers (ISPs) use flat rates to bill their clients. If P2P applications become ubiquitous, they could break the existing business models of many ISPs and force them to change their pricing scheme [9].

Given the considerable traffic volume generated by P2P applications (see our Gnutella estimates in the previous section), it is crucial from the perspective of both their scalability and their impact on the network that they employ available networking resources efficiently. Gnutella’s store-and-forward architecture makes the virtual network topology extremely important. The larger the mismatch between the network infrastructure and the P2P application’s virtual topology, the bigger the “stress” on the infrastructure. In the rest of this section we investigate whether the self-organizing Gnutella network topology maps well to the physical infrastructure.

Let us first present an example to highlight the importance of a “fitting” virtual topology. In Figure 7, eight hosts participate in a Gnutella-like network. We use black, solid, lines to represent the underlying network infrastructure and blue, dotted, lines to denote the application’s virtual topology. In the left picture, the virtual topology closely matches the infrastructure. The distribution of a message generated by node *A* to all other nodes involves only one communication over the physical link *D-E*. In the right picture, the virtual topology, although functionally similar, does not match the infrastructure. In this case, the same distribution involves six communications over the same link.

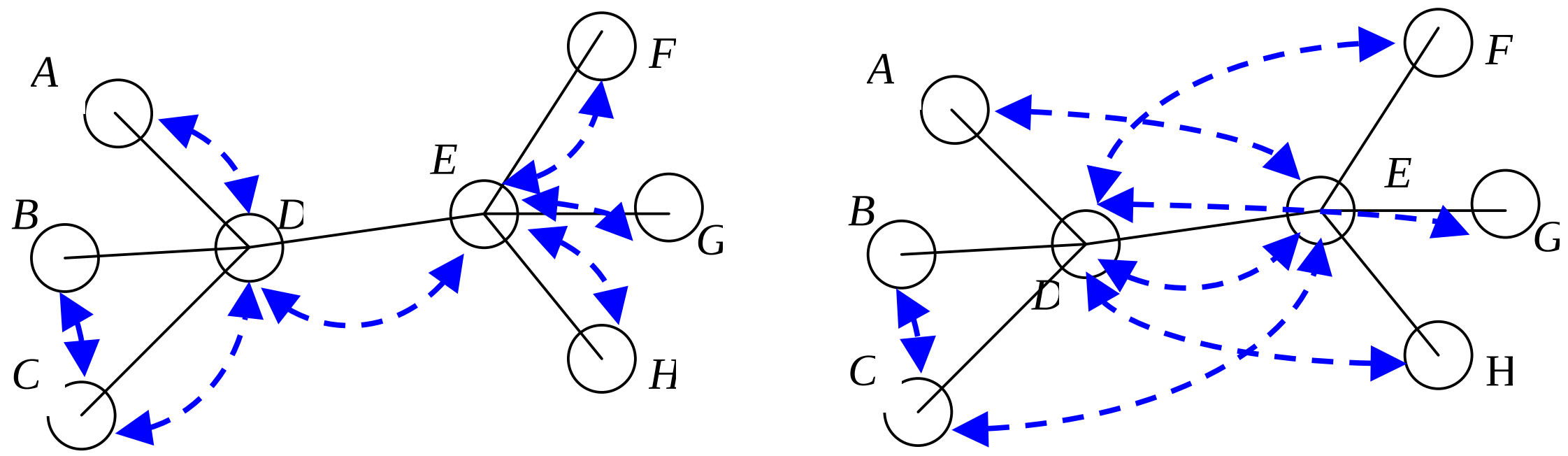

**Figure 7:** Two different mappings of Gnutella's virtual network topology (blue, dotted arrows) to the underlying network infrastructure (black, solid lines). Left picture: perfect mapping. A message inserted into the network by node A travels physical link D-E only once to reach all other nodes. Right picture: inefficient mapping. The same distribution requires that the message traverse physical link D-E six times.

Unfortunately, it is prohibitively expensive to compute exactly the mapping of the Gnutella onto the Internet topology, due both to the inherent difficulty of extracting Internet topology and to the computational scale of the problem. Instead, we proceed with two high-level experiments that highlight the mismatch between the topologies of the two networks.

The Internet is a collection of Autonomous Systems (AS) connected by routers. ASs, in turn, are collections of local area networks under a single technical administration. From an ISP point of view traffic crossing AS borders is more expensive than local traffic. We found that only 2-5% of Gnutella connections link nodes located within the same AS, although more than 40% of these nodes are located within the top ten ASs. This result indicates that most Gnutella-generated traffic crosses AS borders, thus increasing costs, unnecessarily. .

In the second experiment we assume that the hierarchical organization of domain names mirrors that of the Internet infrastructure. For example, it is likely that communication costs between two hosts in the "uchicago.edu" domain are significantly smaller than between "uchicago.edu" and "sdsc.edu." The underlying assumption here is that domain names express some sort of organizational hierarchy and that organizations tend to build networks that exploit locality within that hierarchy.

In order to study how well the Gnutella virtual topology maps on to the Internet partitioning as defined by domain names, we divide the Gnutella virtual topology graph into *clusters*, i.e., subgraphs with high interior connectivity. Given the flooding-like routing algorithm used by Gnutella, it is within these clusters that most load is generated. We are therefore interested to see how well these clusters map on the partitioning defined by the domain naming scheme.

We use a simple clustering algorithm based on the connectivity distribution described earlier: we define as clusters subgraphs formed by one hub with its adjacent nodes. If two clusters have more than 25% nodes in common, we merge them. After the clustering is done, we (1) assign nodes that are included in more than one cluster only to the largest cluster and (2) form a last cluster with nodes that are not included in any other cluster.

We define the entropy [11] of a set C, containing $|C|$ hosts, each labeled with one of the $n$ distinct domain names, as:

$$E(C) = \sum_{i=1}^{n} \left( -p_i \log(p_i) - (1-p_i)\log(1-p_i) \right),$$

where $p_i$ is the probability of randomly picking a host with domain name *i*.

We then define the entropy of a clustering of a graph of size |*C*|, clustered in *k* clusters $C_1, C_2, ..., C_k$ of sizes $|C_1|, |C_2|, ..., |C_k|$, with $|C| = |C_1| + |C_2| + ... + |C_k|$, as:

$$E(C_1, C_2, ...C_k) = \sum_{i=1}^{k} \frac{|C_i|}{|C_1| + |C_2| + ... + |C_k|} * E(C_i)$$

We base our reasoning on the property that $E(C) \geq E(C_1, C_2, ..., C_k)$ no matter how the clusters $C_1, C_2, ..., C_k$ are chosen. If the clustering matches the domain partitioning, then we should find that $E(C) >> E(C_1, C_2, ..., C_k)$. Conversely, if the clustering $C_1, C_2, ..., C_k$ has the same level of randomness as in the initial set *C*, then the entropy should remain largely unchanged. Essentially, the entropy function is used here to measure how well the two partitions applied to set nodes match: the first partition uses the information contained in domain names, while the second uses the clustering heuristic. Note that a large class of data mining and machine learning algorithms based on information gains (ID3, C4.5, etc. [17]) use a similar argument to build their decision trees.

We performed this analysis on 10 topology graphs collected during February/March 2001. We detected no significant decrease in entropy after performing the clustering (all decreases were within less than 8% from the initial entropy value). Consequently, we conclude that Gnutella nodes cluster in a way that is completely independent from the Internet structure. Assuming that the Internet domain name structure roughly matches the underlying topology (the cost of sending data within a domain is smaller than that of sending data across domains), we conclude that the self-organizing Gnutella network does not use the underlying physical infrastructure efficiently.

## 5. Summary and Potential Improvements

The social circumstances that have fostered the success of the Gnutella network might change and the network might diminish in size. P2P, however, "is one of those rare ideas that is simply too good to go away" [18]. Despite recent excitement generated by this paradigm and the surprisingly rapid deployment of some P2P applications, there are few quantitative evaluations of P2P systems behavior. The open architecture, achieved scale, and self-organizing structure of the Gnutella network make it an interesting P2P architecture to study. Our measurement and analysis techniques can be used for other P2P systems to enhance general understanding of design tradeoffs.

Our analysis shows that Gnutella node connectivity follows a multi-modal distribution, combining a power law and a quasi-constant distribution. This property keeps the network as reliable as a pure power-law network when assuming random node failures, and makes it harder to attack by a malicious adversary.

However, Gnutella takes few precautions to ward off potential attacks. For example, the network topology information that we obtain here is easy to obtain and would permit highly efficient denial-of-service attacks. Some form of security mechanisms that would prevent an intruder to gather topology information appears essential for the long-term survival of the network.

We have estimated that, as of June 2001, the network generates about 330 TB/month simply to remain connected and to broadcast user queries. This traffic volume represents a significant fraction of the total Internet traffic and makes the future growth of Gnutella network particularly dependent on efficient network usage. We have also documented the topology mismatch between the self-organized, application-level Gnutella network and the underlying physical networking infrastructure. We believe this mismatch has major implications for the scalability of the Internet—or, equivalently, for ISP business models. This problem must be solved if Gnutella or similarly built systems are to reach larger deployment.

We see two other directions for improvement. First, as argued in [19], efficient P2P designs should exploit particular distributions of query values and locality in user interests. Various Gnutella studies show that the distribution of Gnutella queries is similar to the distribution of HTTP requests in the Internet: they both follow a Zipf's law (note that, although the Zipf's formulation is widely used, these distributions can also be expressed as power-law distributions). Therefore, the proxy cache mechanism used in the Web context might have useful applications in a P2P context. Moreover, when nodes in a dynamic P2P network are grouped by user interest, a query-caching scheme could bring even larger performance improvements.

A second direction of improvement is the replacement of query flooding mechanism with smarter (less expensive in terms of communication costs) routing and/or group communication mechanisms. Several P2P schemes proposed recently fall into the former category: systems like CAN [20] or Tapestry [21] propose a structured application-level topology that allows semantic query routing. We believe, however, that a promising approach is to preserve and benefit from the power-low characteristics that, as shown in this paper, emerge in Gnutella's ad-hoc network topology. A way to preserve the dynamic, adaptive character of the Gnutella network and still decrease resource (network bandwidth) consumption is to use dissemination schemes (e.g., based on epidemic protocols) mixed with random query forwarding. We have collected a large amount of data on the environment in which Gnutella operates, and plan to use this data in simulation studies of protocol alternatives.

## 6. Acknowledgements

We are grateful to Larry Lidz, Conor McGrath, Dustin Mitchell, Yugo Nakai, Alain Roy, and Xuehai Zhang for their insightful comments and generous support. This research was supported in part by the National Science Foundation under contract ITR-0086044.